\documentclass[aps, superscriptaddress, showpacs, floatfix, twocolumn]{revtex4-2}
\usepackage{graphicx}
\usepackage{hyperref}
\usepackage{amsmath}
\usepackage{color}
\usepackage{graphicx}
\begin{document}
\title{A model for cooperative scientific research inspired by the ant colony algorithm}
\author{Zhuoran He}
\affiliation{School of Physics, Huazhong University of Science and Technology, Wuhan 430074, China}
\author{Tingtao Zhou}
\affiliation{Division of Engineering and Applied Sciences, California Institute of Technology, Pasadena, CA 91125, USA}

\begin{abstract}
\vspace*{0\baselineskip}
Modern scientific research has become largely a cooperative activity in the Internet age. We build a simulation model to understand the population-level creativity based on the heuristic ant colony algorithm. Each researcher has two heuristic parameters characterizing the goodness of his own judgments and his trust on literature. In a population with all kinds of researchers, we find that as the problem scale increases, the contributor distribution significantly shifts from the independent regime of relying on one's own judgments to the cooperative regime of more closely following the literature. The distribution also changes with the stage of the research problem and the computing power available. Our work provides some preliminary understanding and guidance for the dynamical process of cooperative scientific research in various disciplines.
\end{abstract}

\maketitle

\section{Introduction}
Cooperative scientific research is a new trend in the science community nowadays due to the growth of number of researchers \cite{bloom2020ideas,larson2014too,engineeringbythenumbers}, the faster propagation of knowledge through the Internet \cite{ginsparg2011arxiv,goodrum2001scholarly,holmberg2014disciplinary,hurd2000transformation} and the many new interdisciplinary research topics \cite{Yin2017,Bazzani2009}, etc. Research groups ranging from a few scientists to international institutions can study related problems and build upon each other's works. In the early pioneering days, the activity of scientific research was the solitary work of a few geniuses of the world and the spirits of independent thinking and skepticism were highly valued. In modern days, we are seeing more and more scientific achievements made by the progressive efforts of many researchers. 
This paradigm shift accompanies the development of complexity science itself \cite{Yin2017,Bazzani2009,anderson1972more}. Scientists in the Internet age work like a highly cooperative ant colony connected by pheromone, i.e., research publications, and exhibit population-level creativity which requires modeling to understand and optimize.

Previous studies on scientific research and collective intelligence have discussed various aspects of this topic including the citation system \cite{garfield1955citation,leydesdorff1998theories,hirsch2007does,fong2017authorship},  evaluation and funding system~\cite{hicks2012performance,muscio2013does}, game theory competition and cooperation~\cite{lim2018satisfied,tiokhin20competition,sonubi2016effects,axelrod1997complexity}, complex networks~\cite{barrat2004architecture,newman2001clustering}, team size and composition management~\cite{milojevic2014principles,massey2006crossing,wu2019large}, and so on~\cite{goldman1991economic,kealey1996economic,wu2015impact,wakeling2001intelligent,karamched2020heterogeneity,durve2020collective}. 
In this paper, we build a simplified model inspired by the ant colony optimization (ACO) algorithm \cite{dorigo1992optimization,dorigo1991distributed,dorigo1996ant} to study the dynamical process of cooperative scientific research by computer simulations. Our ant colony model can obtain the optimal research styles for various types of scientific problems, e.g., simple (elemental) v.s. complex, new v.s.~old (long-standing), etc., and study the influence of computing power and different survival rules on selecting researchers for the community.

We suppose that in the ant community, each scientific problem they study is a randomly generated traveling salesman problem (TSP)~\cite{flood1956traveling} with $N$ vertices, where $N$ controls the complexity of the problem. A researcher's effort on such a problem is modeled as making small decisions step by step to connect the vertices and find a plausible path. He will then pass on the knowledge by publications, i.e., leaving pheromone on the edges visited. The shorter the total path, the more pheromone will be assigned. Every decision made is governed by two heuristic parameters:  $\alpha$ characterizing the researcher's trust on published literature, i.e., the pheromone left on all edges, and $\beta$ characterizing his trust on the greedy local distance measure, i.e., the researcher's own sense of direction. The procedure is iterated as generations of researchers attempt for better solutions. Finally, the accumulated pheromone concentrates on the shortest TSP path found by the community, which represents the currently best answer known to the scientific problem.

Two essential ingredients of our ant colony model are the NP-hardness of TSP and the pheromone mechanism in ACO. Since TSP is NP-hard, it is easy to evaluate and compare path lengths and exclude the longer path as `wrong', but difficult to know if the shorter path is indeed shortest, which is similar to open questions in science that satisfy the falsifiability criterion. The pheromone is a population-level information sharing mechanism that enables researchers to work out difficult scientific problems cooperatively. Our ant colony model develops the ACO in that we have improved the pheromone update rules and allow the heuristic parameters $\alpha,\beta$ to differ individually and change by evolution. We can then study the equilibrium distributions of $\alpha,\beta$ given different problem scales $N$ and different numbers of ACO iterations that distinguish between new and old problems. The influence of computing power will be modeled by introducing the Hamiltonian cycle speedup \cite{Croes1958} that mimics the role of computers.

\section{Ant colony model}
\subsection{The core ACO algorithm}
In the ACO algorithm, each ant with two heuristic parameters $\alpha,\beta$ tries to find a TSP path individually. The ant has its own memory of the set of vertices $S$ that has been visited and has access to the community-shared information $d_{ij}$, the distance between vertices $i,j$ and $\tau_{ij}$, the amount of pheromone on the undirectioned edge $i-j$. The ant picks a random vertex to start the trip. Then each step from vertex $i$ to vertex $j$ is determined by the transition probability
\begin{equation}
P_{i\rightarrow j} \propto\,\begin{cases}
\;(0.01+\tau_{ij})^\alpha\,/\,d_{ij}^\beta, & j\not\in S, \\
\;0, & j\in S.
\end{cases}
\label{eq:ACO-SAW}
\end{equation}
The probabilities $P_{i\rightarrow j}$ are normalized for all $j\not\in S$ to determine the next stop $j$. Then vertex $j$ is added to set $S$ so it will not be repeatedly visited. Eq.~\eqref{eq:ACO-SAW} describes the basic rules of the heuristically biased self-avoiding walk (SAW) \cite{slade2019selfavoiding, LI2020105257} of ants in the original ACO algorithm. We have added a small background value of $0.01$ to $\tau_{ij}$ in Eq.~\eqref{eq:ACO-SAW} so that the ants do not get oversensitive to small amounts of pheromone.

We have also made improvements in the pheromone update rules. After $N_\mathrm{ants}=50$ ants have finished their TSP paths, we pick the best $p(t)$ percent and allow these winning ants to leave pheromone over their TSP paths bidirectionally. The amount of pheromone on each edge is inversely proportional to the total path length and proportional to a linearly decreasing weight of the path ranking. Long-distance steps on the TSP path gets extra penalties. The pheromone on all edges then evaporates by $p(t)$ percent and the above procedure iterates while the percentage $p(t)$ gradually decreases from 50\% to 8\% (4~ants) over the iterations.

These improvements mean that initially the ant colony is very eager to accumulate pheromone and later, the update rules get tighter as the best-known path of the ant colony becomes nearly optimal. But any time, an ant who beats the best-known path always immediately becomes the leader of the top 4 ants and leaves the most pheromone to the whole ant colony. With the improved pheromone update rules, the pheromone becomes a more useful guide to the ants and better resembles the literature publication system in academia. More details of the model can be found in our Matlab code provided upon reasonable request.

\subsection{Evolution of heuristic parameters}
In the original ACO algorithm, the parameters $\alpha,\beta$ were set manually as hyper-parameters and applied to all ants. In our model, we allow $\alpha,\beta$ to take different values for different ants and evolve the distribution $P_{\alpha,\beta}$ by training the ant colony with randomly generated TSP graphs. During the solution of one graph, $P_{\alpha,\beta}$ is kept unchanged and the heuristic parameters of the ants who found shorter TSP paths than the best-known path are recorded. These ants are called \textit{contributors} and their $\alpha,\beta$ values are used for evolving $P_{\alpha,\beta}$ according to
\begin{equation}
P_{\alpha,\beta}^\mathrm{(new)} = \frac{n_c}{M}P_{\alpha,\beta}^{(c)} + \left(1-\frac{n_c}{M}\right) P_{\alpha,\beta}^\mathrm{(old)}.
\label{eq:evolution}
\end{equation}
Here $n_c$ is the number of contributors recorded during one graph and $M=4000$ is the total pool of ants out of which the $N_\mathrm{ants}=50$ ants are sampled in each ACO iteration. When equilibrium is reached, every ant in the colony is equally likely to become a contributor. More favorable $(\alpha,\beta)$ values will attract more ants and less favorable values will be adopted by fewer ants.

We then consider a more sophisticated situation where the trained distribution $P_{\alpha,\beta}(t)$ can depend on problem stage $t$. To do this, we record for each contributor not only its $\alpha,\beta$ values, but also the number of ACO iterations $t$ performed when its contribution is made. We can then compare at equilibrium the distributions $P_{\alpha,\beta}(t)$ suitable for different problem stages $t$.

\subsection{Hamiltonian cycle speedup}
The Hamiltonian cycle speedup \cite{Croes1958} is often used in conjunction with ACO to speed up its convergence. In the core ACO algorithm, at every vertex $i$, the ant heuristically picks a next vertex $j$ based on $P_{i\rightarrow j}$, which mimics human intuition. The TSP path obtained is called a Hamiltonian cycle in graph theory, e.g.,
\begin{equation}
1\rightarrow (i-1)\rightarrow i \rightarrow\cdots\rightarrow j\rightarrow (j+1)\rightarrow N \rightarrow 1.
\label{eq:Hamiltonian-cycle}
\end{equation}
The Hamiltonian cycle speedup plays the role of a computer exhaustively checking human errors. It enumerates all segments $i\rightarrow\cdots\rightarrow j$ of the Hamiltonian cycle in Eq.~\eqref{eq:Hamiltonian-cycle} and checks if the cycle length can be made shorter by reversing the segment into $j\rightarrow\cdots\rightarrow i$, which is true if and only if $d_{i-1,i}+d_{j,j+1}>d_{i-1,j}+d_{i,j+1}$. The exhaustive check continues until no such improvements are possible, which is a necessary but not sufficient condition for the optimality of the TSP path. We examine the influence of the Hamiltonian cycle speedup on the distribution $P_{\alpha,\beta}$ in Sec.~III$\,$C and use the simpler model described in Secs.~II$\,$A--II$\,$B elsewhere.

\section{Results and discussions}
\subsection{Effect of problem scale}
\begin{figure}
\includegraphics[width=\columnwidth]{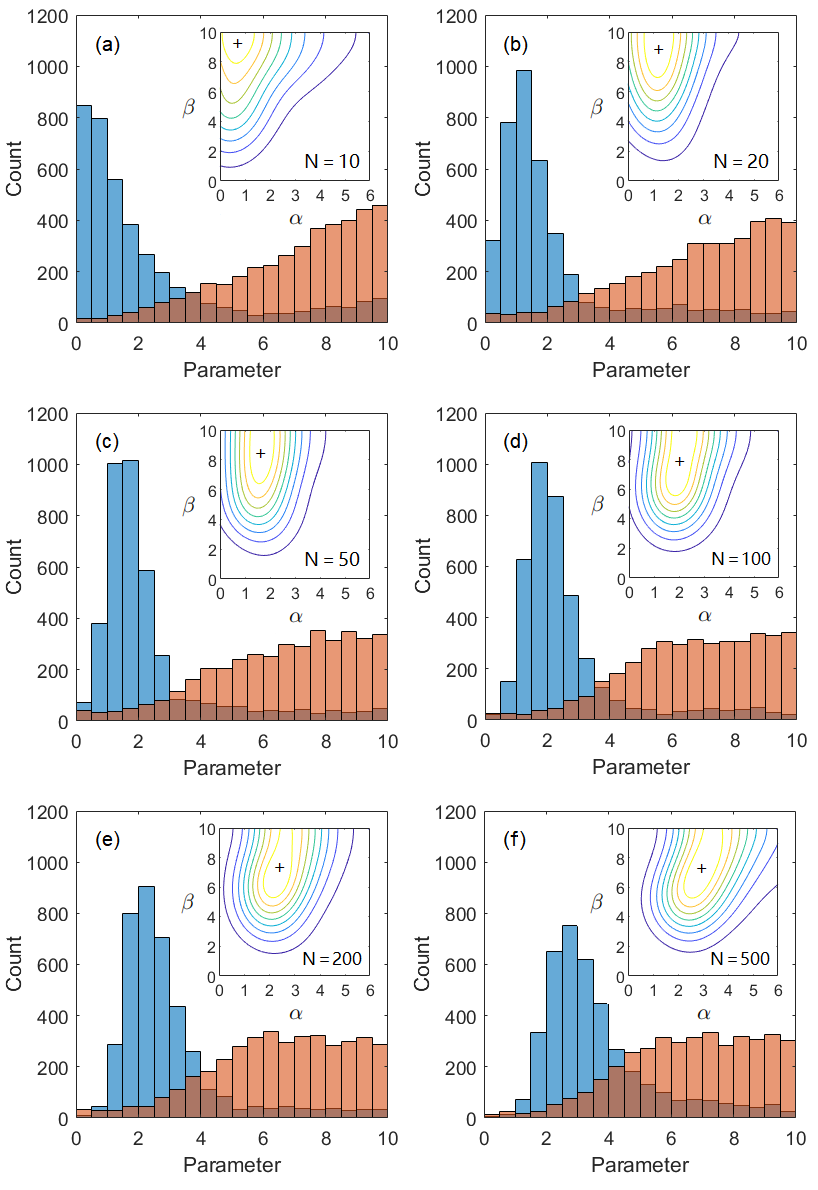}
\caption{Histograms of heuristic parameters $\alpha$ (blue) and $\beta$ (red) and contour plots of their joint distributions (insets) for TSP graphs with $N = 10$, 20, 50, 100, 200, and $500$ vertices. The ``+'' sign marks the mode peak.
\label{fig:problem-size}}
\end{figure}
We first examine how the equilibrium distribution $P_{\alpha,\beta}$ changes with problem scale $N$, i.e., the number of vertices in the TSP graph. The vertices are randomly sampled from a uniform distribution in a 2D unit square region. We have tried other distributions (Gaussian, triangular) and other region shapes (rectangle, circle) and have found qualitatively the same results.
 
As shown in Figs.~\ref{fig:problem-size}a--\ref{fig:problem-size}f, the $\alpha$ peak significantly shifts to larger values as the problem scale $N$ increases. This indicates that when faced with more difficult problems, the research community has to rely more on previous works for guidance to find sophisticated better solutions, and random trials ignoring literature is not as efficient. The $\beta$ parameter governs the researcher's goodness of local distance measure or sense of direction. For small problems, the optimal distribution relies heavily on high $\beta$ values. For larger problems, the weights of high $\beta$ reach a plateau and the joint distribution $P_{\alpha,\beta}$ develops a positive correlation between $\alpha$ and $\beta$, suggesting that the successful research style is a combination of the $\alpha$ and $\beta$ heuristics. Such researchers always keeps up-to-date knowledge of the currently best solution known by the community ($\alpha$ heuristics) and quickly identifies where potential improvements are possible ($\beta$ heuristics) around the community-found path.

\subsection{Time-dependent distribution}
\begin{figure}
\includegraphics[width=\columnwidth]{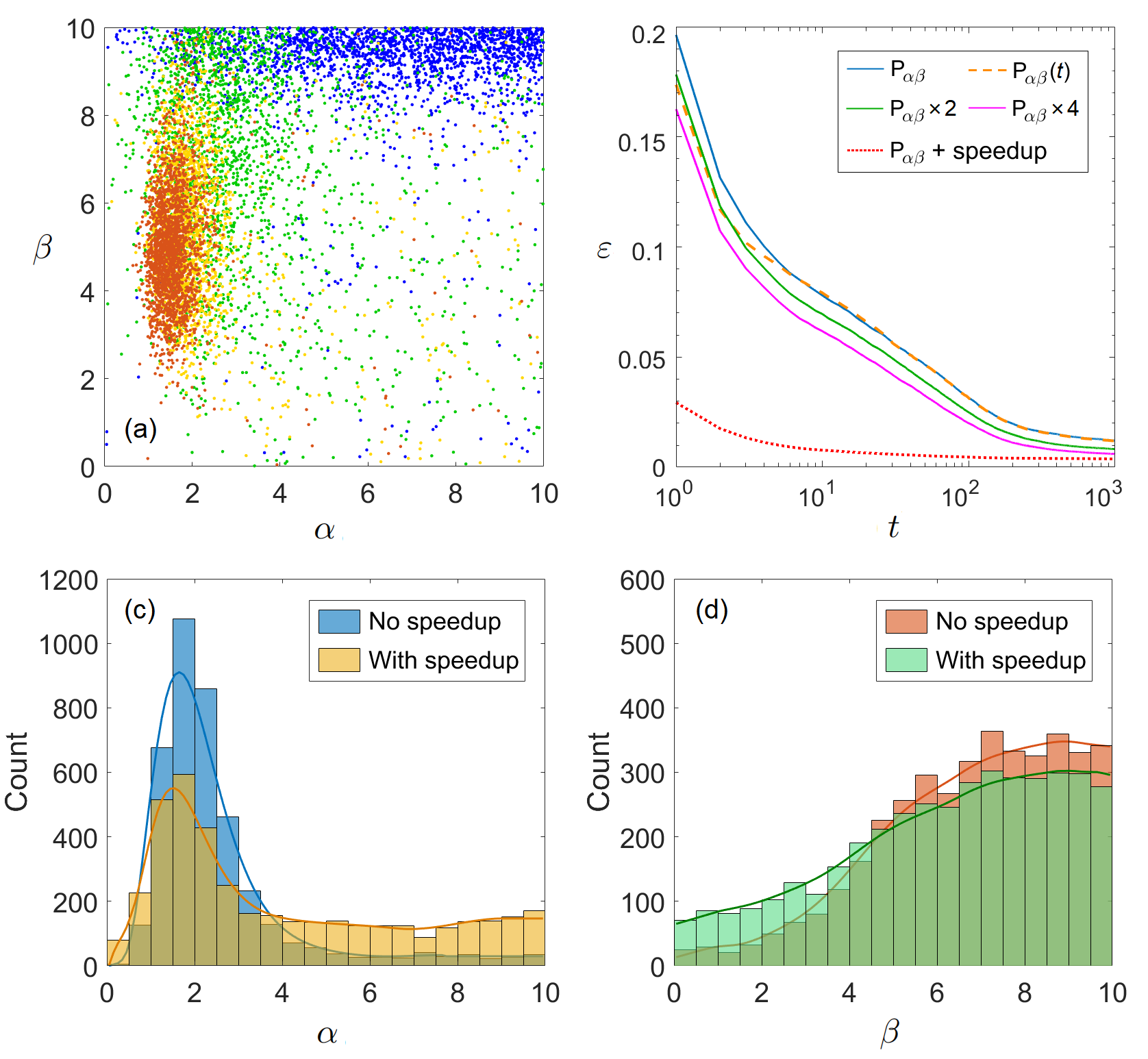}
\caption{(a) The $P_{\alpha,\beta}(t)$ distribution with $t=1$--$5$, $6$--$30$, $31$--$100$ and $101$--$1000$ scatter plotted in blue to red. (b) Error curves of $P_{\alpha,\beta}(t)$ compared with $P_{\alpha,\beta}$ and other variations. Panels (c)--(d) compare the $P_{\alpha,\beta}$ distributions with and without the Hamiltonian cycle speedup. Number of vertices $N=100$ in all $4$ subplots.
\label{fig:problem-stage}}
\end{figure}
Some difficult problems can persist for years or decades as researchers come and leave. In an ideal situation, researchers switching from problems to problems specialize to both the appropriate scale of complexity and the stage of problem conducive to their own research styles ($\alpha,\beta$ values). We therefore consider $P_{\alpha,\beta}(t)$ with $t$ being the number of ACO iterations for fixed problem scale $N=100$. We train $P_{\alpha,\beta}(t)$ to equilibrium and plot the results in Fig.~\ref{fig:problem-stage}a in 4 colors corresponding to 4 problem stages: newly proposed ($t=1$--$5$, blue), early ($t=6$--$30$, green), intermediate ($t=31$--$100$, yellow), and late ($t=101$--$1000$, red) periods.

When a problem is newly proposed, the contributors (blue) generally have high $\beta$ values. Since there are not many publications to read yet, researchers with low $\beta$ will move randomly between the vertices and obtain TSP paths of order $\mathcal{O}(N)$, while those with high $\beta$ will always greedily choose the closest vertex to move to. The greedy solution can be estimated to be
\begin{equation}
\mathcal{O}\left(\frac{1}{\sqrt{N}}+\frac{1}{\sqrt{N-1}}+\cdots+1\right)=\mathcal{O}(\sqrt{N}),
\end{equation}
which is much better than a random self-avoiding walk $\mathcal{O}(N)$. Therefore, all contributors of a newly proposed problem tend to have high $\beta$ values. After the greedy solution has been found, the early-stage contributors (green) constitute the upper part of the time-independent distribution $P_{\alpha,\beta}$ in Fig.~\ref{fig:problem-size}d. The intermediate (yellow) and late-stage (red) contributors then scan down to the lower part of $P_{\alpha,\beta}$ and finally concentrate into a red blob below the mode peak of $P_{\alpha,\beta}$.

The error curves of the time-dependent $P_{\alpha,\beta}(t)$ and time-independent $P_{\alpha,\beta}$ distributions are compared in Fig.~\ref{fig:problem-stage}b. We use the relative error $\varepsilon(t)=L(t)/L_\mathrm{opt}-1$ averaged over 500 graphs to evaluate the goodness of a given research condition, where $L_\mathrm{opt}$ is the optimal path length obtained from the open-source exact TSP solver {\it Concorde}~\cite{applegate2006concorde,applegate2009certification}. Initially, the greedy solution of $P_{\alpha,\beta}(t)$ (yellow dashed line in Fig.~\ref{fig:problem-stage}b) has an advantage over $P_{\alpha,\beta}$ (blue line in Fig.~\ref{fig:problem-stage}b), which does not last for very long. The blue and yellow curves nearly coincide when the problem reaches intermediate to late periods. The residue error at $t=10^3$ remains $\sim 1.2\%$, which is likely to result from the path dependence effect \cite{Paul.David1985}, i.e., the ant colony gets trapped to a local minimum found by previous works. If we have two or more independent research communities (green \& purple lines in Fig.~\ref{fig:problem-stage}b), which is realized by running the ant colony code multiple times and keeping the smallest TSP length of the trials at every iteration step $t$, the relative error $\varepsilon(t)$ has a statistically significant reduction.

\subsection{Effects of computing power}
We then move on to discuss the effects of more computing power, which is mimicked by introducing the Hamiltonian cycle speedup described in Sec.~II$\,$C. When individual researchers have computers that help them do exhaustive trials and verifications, our results indicate that the selectivity effects on both the literature parameter $\alpha$ and the intuition parameter $\beta$ of the contributors are significantly reduced. As is shown in Figs.~\ref{fig:problem-stage}c-\ref{fig:problem-stage}d, both the $\alpha$ peak and the $\beta$ plateaus are made lower by introducing the Hamiltonian cycle speedup. This means that computers are a chance equalizer which diversifies the heuristic parameter distributions of the contributors. Also, the $\alpha$ peak slightly shifts to smaller values, which is due to the reduction of effective problem hardness when computers become available. In terms of relative error (red line in Fig.~\ref{fig:problem-stage}b), the introduction of Hamiltonian cycle speedup significantly speeds up the convergence to the optimal TSP path. The residue error at $t=10^3$ iterations $\sim 0.4\%$ is made much smaller than the blue curve but still nonzero, which suggests that the path dependence effect of ant-colony research cannot be completely eliminated even with more computing power available to each researcher individually.

\subsection{Other non-ideal situations}
\begin{figure}
\includegraphics[width=\columnwidth]{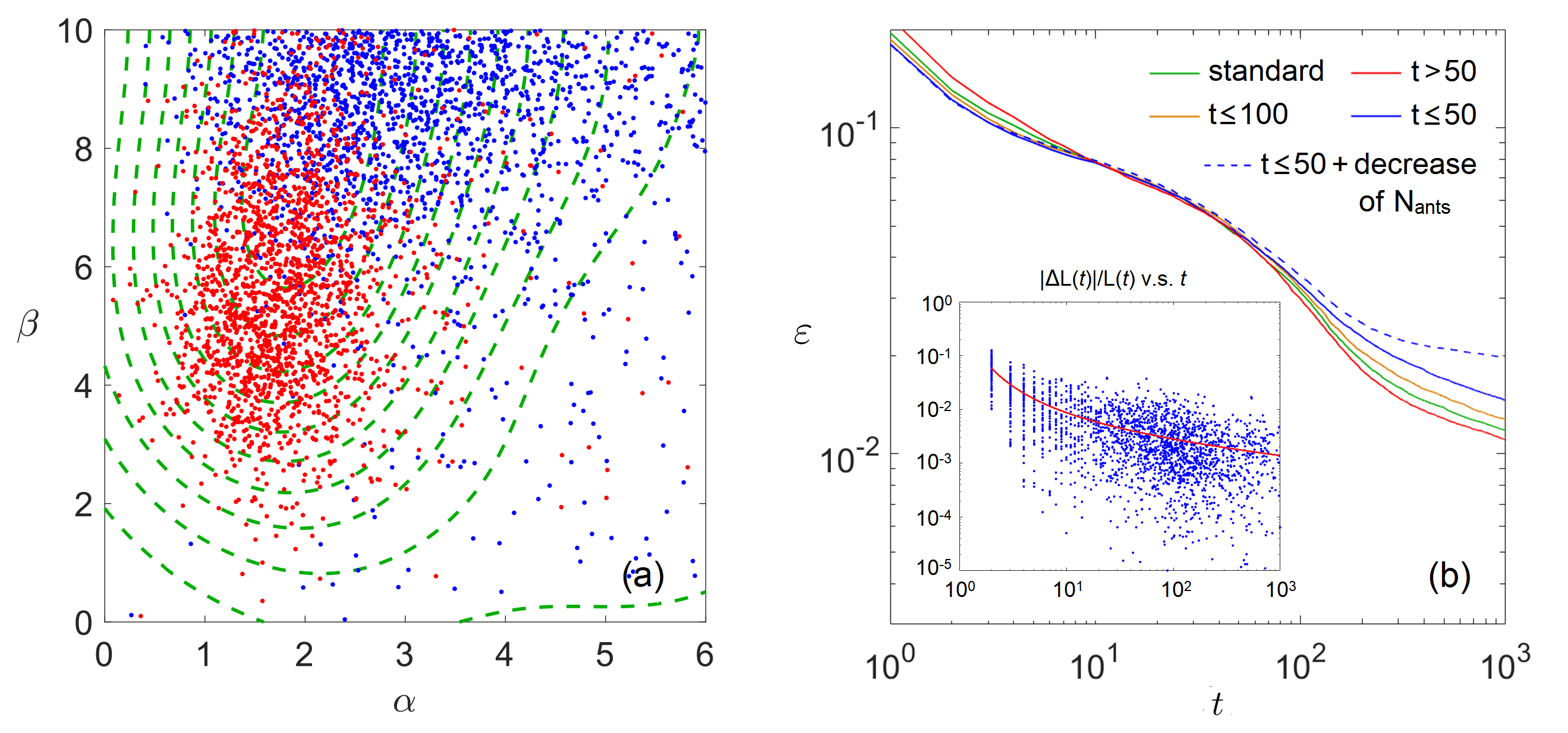}
\caption{(a) The equilibrium contributor distribution under the survival rules of $t\leq 50$ (blue) and $t>50$ (red). Dashed contours are those of the standard $N=100$ distribution in Fig.~\ref{fig:problem-size}d. (b) The error curves $\varepsilon(t)$ of different survival rules. Inset shows normally trained contributor distribution in terms of improvement percentage v.s. problem stage.
\label{fig:early-stopping}}
\end{figure}

We often see in academia that researchers are faced with tight and pressing survival rules, most of which are achievement-based. We find in our model that sometimes these rules can be counter-productive to the science community. An important reason why this happens is that such rules would encourage researchers to focus on new or early-stage problems, leaving the late-stage problems simply ``outdated'' rather than solved.

We simulate such a situation and results are shown in Fig.~\ref{fig:early-stopping}. Suppose a problem is interesting to the ant colony for only $t\leq 50$ iterations, after which the problem becomes old and out of attention. By training the ant colony using TSP graphs with $N=100$ vertices under such hasty rules, the equilibrium distribution $P_{\alpha\beta}$ is given by the blue dots in Fig.~\ref{fig:early-stopping}a. Conversely, if every graph is solved up to $t=10^3$ iterations but only those contributors after $t>50$ are recorded to update $P_{\alpha\beta}$, the ant colony will be trained into the distribution of the red dots. The green contours in Fig.~\ref{fig:early-stopping}a show the normally trained distribution where all contributors are recorded to update $P_{\alpha\beta}$. Since achievement-based survival rules pick out those contributors with big improvements of TSP lengths, which, according to the inset of Fig.~\ref{fig:early-stopping}b, tend to be early-stage contributors, the distribution $P_{\alpha\beta}$ shifts to the blue side as a result.

We then plot in Fig.~\ref{fig:early-stopping}b the error curves $\varepsilon(t)$ of different $P_{\alpha\beta}$ distributions averaged over 500 graphs. The blue distribution has short-term benefits but long-term costs. A ``hasty'' ant colony adapted to early-stage problems would lack those ants with heuristic parameters conducive to making breakthroughs on long-standing problems and therefore become inefficient as problems approach late stages. In reality, the combined effect of making the researcher community both inefficient and not interested in solving long-standing problems could be even worse, which can be mimicked by reducing $N_\mathrm{ants}=50\,e^{-(t-50)/200}$ after $t>50$ (blue dashed line in Fig.~\ref{fig:early-stopping}b). The residue error at $t=10^3$ reaches $\sim 2\%$. More interestingly, the normally trained $N=100$ distribution (green line in Fig.~\ref{fig:early-stopping}b) can be outperformed by the red distribution (red line in Fig.~\ref{fig:early-stopping}b) in the long run, which suggests the importance of giving more weights to the late contributors.

\section{Conclusion}
We have established an ant-colony research model which enables us to understand the dynamical process of cooperative scientific research in various disciplines. Based on our model, we have made several interesting findings. First, as the problem scale increases, the contributors tend to have more cooperative heuristic parameters than those of simpler problems. Therefore, the cooperative mode of scientific research is a consequence of complexity science itself. Second, different problem stages will require different research styles. In the beginning, simple intuitive thinking can help lay down the general framework. Later, improvements become harder and require deeper thinking and more trials and errors. Third, the introduction of computers or any other advanced technology that enables exhaustive trials and verifications can give the human researcher more freedom, diversify the contributor population and make the scientific results more accurate and objective.

In addition to demonstrating the power of cooperative scientific research, our model can also simulate non-ideal situations and identify how things might go wrong. First is path dependence. As scientists build upon each other's works, there is inevitably some degree of path dependence. Parallel development of several independent communities, technological methods, or schools of thoughts can be better than having one unified community stuck with pre-established ideas and paradigms. Second is hasty research. Putting pressure on productivity or individual achievements can lead to hasty research. It is important to give more credits to late contributors and solvers of long-standing problems for the long-term progress of science.





\section*{Acknowledgments}
We thank the helpful discussions with colleagues in Department of physics at Caltech. T.~Z. acknowledges support by the Cecil and Sally Drinkward Fellowship.

\bibliography{aco}
\end{document}